# Visualization of Retrieved Documents using a Presentation Server

Sa-Kwang Song & Sung Hyon Myaeng

Department of Computer Science

Chungnam National University, Taejon, Korea

sksong@cs.chungnam.ac.kr, shmyaeng@cs.chungnam.ac.kr

## Abstract

In any search-based digital library (DL) systems dealing with a non-trivial number of documents, users are often required to go through a long list of short document descriptions in order to identify what they are looking for. To tackle the problem, a variety of document organization algorithms and/or visualization techniques have been used to guide users in selecting relevant documents. Since these techniques require heavy computations, however, we developed a presentation server designed to serve as an intermediary between retrieval servers and clients equipped with a visualization interface. In addition, we designed our own visual interface by which users can view a set of documents from different perspectives through layers of document maps. We finally ran experiments to show that the visual interface, in conjunction with the presentation server, indeed helps users in selecting relevant documents from the retrieval results.

# 1. Introduction

In any search-based digital library (DL) systems dealing with a non-trivial number of documents, users are often required to go through a long list of retrieved document descriptions in order to identify what they are looking for. Since the users are not fully aware of the internal workings of the retrieval engine embedded in the DL systems, they are forced to sequentially bring in each document (the whole or a significant portion) and check for its relevance until they are satisfied or out of time. This time-consuming process is not likely to be avoidable in the foreseeable future because of the inherent difficulty and inaccuracy of information retrieval (IR). The problem can get worse, even with the state-of-art IR technologies, as more and more information is going to be available through digital libraries.

As a way of alleviating this problem, researchers have devised ways to allow users to visualize queries and/or retrieval results in a two or three dimensional space [4, 5, 8, 10]. A variety of visualization techniques such as fish-eye views [15], cone-trees [13], or even virtual reality techniques [1] have been used to represent information objects users can manipulate. Underlying these user interfaces are the methods for organizing objects based on their characteristics and similarities (e.g. [3, 8, 14]). These two groups of techniques together are supposed to help users navigate through the space of objects in their own ways and reduce human efforts in obtaining the desired information.

Continuing and building on the two lines of research efforts, we propose a way of making visualization and information organization more practical in digital library environments. More specifically, we developed a presentation server whose sole purpose is to receive retrieval results from a retrieval server, re-organize them, and send the result to a client computer that has a visualization interface. The presentation server is located between clients and multiple retrieval servers, and works essentially as an intermediary between the client and the retrieval server. The retrieval results provided by the retrieval server are organized and passed to the client where a visualization technique is used.

This type of presentation server can reduce the burden of a client that processes the data in a potentially complicated way and visualize them, thus making it possible to keep the client as light as possible. Furthermore, the server prevents clients from having to directly deal with different formats of retrieval results obtained from different retrieval servers. When the client is implemented by Java applets and plug-ins, which is the case in our work, the visualization module can be dynamically loaded from the presentation server. Incidentally, this implementation scheme makes the idea even more viable because the user can select from a variety of visualization methods available from the presentation server.

In addition, this paper introduces a new way of visualizing retrieval results, which takes advantage of the presentation server. By showing different views of the retrieval results, the user interface in the client allows the user to guess their contents without having to actually read the documents proper. There are several layers of maps, each of which contains 100 cells (10 by 10) that correspond to 100 top-ranked documents and shows a unique view of the documents. By switching back and forth among the layers, the users can see different facets of the group of the document. In our current implementation, some layers indicate to what extent each document contains a particular query term, and others show whether documents belong to different clusters.

We first show in Section 2 the architecture of the presentation server, together with the other components in our digital library system, and discuss some implementation issues. In Section 3, we describe the user interface in the client, which receives the data generated by the presentation server and visualize them to help users reduce their cognitive efforts in selecting the desired documents. In order to validate our design and show its utility, we ran some experiments whose results indeed demonstrate that users can benefit from multiple views of the retrieved documents. These results, which would not have been obtainable without a presentation server, are described in Section 4. The last section gives our conclusions and future work.

## 2. The Architecture for the Presentation Server

In a typical retrieval environment, a user query is sent to the retrieval server that processes the query and sends back a list of document surrogates such as short titles. The list is usually based on similarity values calculated by the retrieval algorithm in the server. The presentation server architecture we propose in this paper assumes the same kind of retrieval environment, but adds flexibility of providing more sophisticated user interfaces or richer user/system interactions in a practical way. In a digital library environment where multiple retrieval servers may exist and many clients are connected to them, the idea of assigning presentation-related functions to a separate server has a particular importance of making the clients light and reducing the load on the servers.

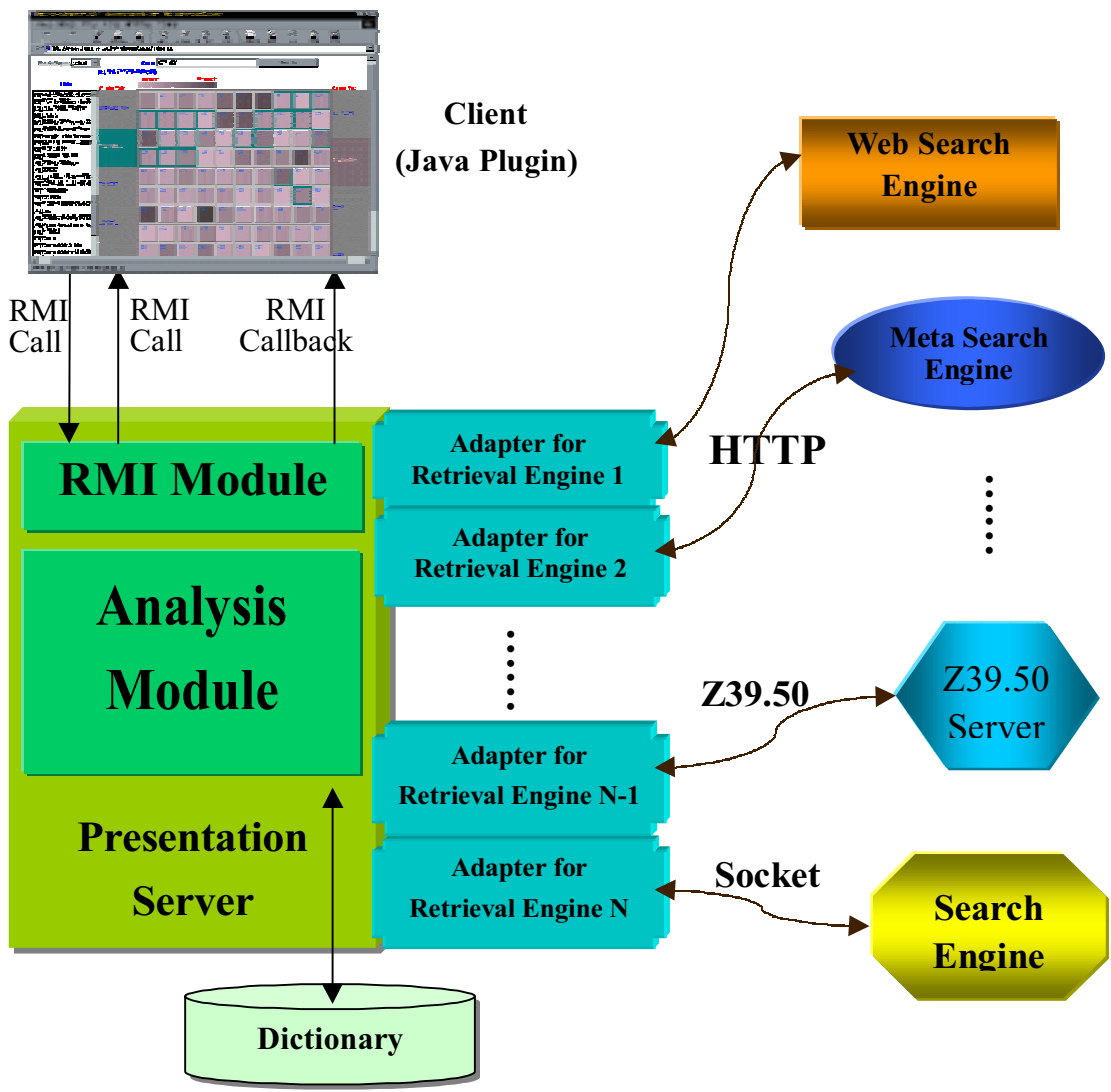

Fig. 1. The architecture for the presentation server

As in Fig. 1, the presentation server is located in between the retrieval servers and the clients. It receives a user's query and bypasses it to an appropriate retrieval server and receives the retrieval result, typically as a list of documents. The main function of the presentation server is to process the information about the retrieved documents, which is available in the initial list or obtained by making a request for additional information to the retrieval server, and send the processed information to the visualization module in the client.

The presentation server consists of two main modules: the *RMI (Remote Method Invocation) module* in charge of communications with the client and the *analysis module* responsible for all computational work in preparation for visualization and presentation. The RMI module handles connection requests and queries from the client and replays the document information produced by the analysis module to the client, whereas the analysis module does all the heavy computations such as clustering or text analysis for the retrieved documents.

The RMI module is responsible for understanding the client requests by means of Java RMI communication methods. RMI provides for a mechanism by which an object in one Java virtual machine can call a method of an object in another Java virtual machine [12]. The RMI module in our server receives a request from the client and activates a thread that is exclusively responsible for handling the particular request. This mechanism not only allows for a stateful connection between the client and server, unlike the HTTP protocol, but also reduces unnecessary overhead associated with the creation of threads and allocation of memory. Furthermore, the current implementation uses mutual RMI call methods for efficiency rather than the basic RMI calls [12].

The analysis module can be implemented in a number of different ways, depending on the data required by the visualization methods in the client. For example, if document objects are plotted in a three dimensional space, similarities among all the documents to be displayed as objects need to be computed. Since the usual document surrogates such as titles are not descriptive enough, their full descriptions need to be obtained from the retrieval engine. In our current implementation where clustering and re-ranking of documents are needed, the analysis module includes a typical indexing procedure and a clustering method.

The client is implemented with Java applets and exchanges data with the presentation server by means of RMI calls. In order to ensure that the client functions are performed regardless of the browser software, we use Java plug-in's that enables web page authors to direct Java applets or JavaBeans components on their web pages to run using Java Runtime Environment (JRE), instead of the browser's default runtime [7]. In this way, we can apply the state-of-art Java technology to the applets and obtain the same functionality regardless of the types of the client browser.

The *adapters* are capable of interpreting the meaning of the retrieval result coming from the corresponding retrieval server and converting this data and other additional document information to the format that the analysis module can process. As illustrated in Fig. 1, the adapters also serve as a communication interface to retrieval engines that understand different communication protocols. For example, some adapters built for Internet search engines use Java URL class methods to extract HTML documents from the data stream, whereas others built for Z39.50 retrieval engines need to have Z39.50 client functionality. Since the Z39.50 client functionality is no longer necessary in the user client, we achieve the goal of making it light.

When the adapters for a variety of retrieval engines are provided, users can even switch back and forth between different retrieval engines regardless of the communication protocols and compare the results dynamically. This capability also opens up the possibility of combining multiple retrieval results in the presentation server and sending a single list to the client.

## 3. Visual Display of Multiple Views

The main thrust of our visual display method is to show multiple views of retrieved documents so that users can quickly deduce their contents without having to actually bring in the full document descriptions. Since a set of documents can be displayed graphically in many different ways, each perhaps with certain distortions due to the limited information that can fit in the limited display area, we attempt to show their different perspectives on a display almost at the same time.

In our current design and implementation, different perspectives are in two types: document rankings based on different combinations of query words and document clustering results. The visual interface allows the users to easily compare ranking results with respect to different query terms or sub-queries and the clustering result for the entire set of the retrieved documents. Our assumption is that with multi-way comparisons, it would be easier for the users to predict the relevance of the documents and choose to read the details of those that are more likely to satisfy them than others. This flexibility is expected to help users resolve the discrepancy between the way retrieval engines rank the documents and the way the users would expect them to be ranked, which occurs due to the lack of understanding the internal workings of retrieval models, typically Boolean or vector space models. This idea is in line with the visualization technique used in text tiling where pages of long documents are shaded differently for the degree of their relevancy with respect to two query terms [4].

Document clustering has been an important technique by which retrieval results can be organized before they are presented to the user so that they can narrow down the search space quickly[2]. Our contribution, however, is not on development of a new clustering algorithm but on the way clustering results are displayed and used in conjunction with the multiple rankings. More specifically, we provide a clustering result together with the similarity values assigned to individual documents.

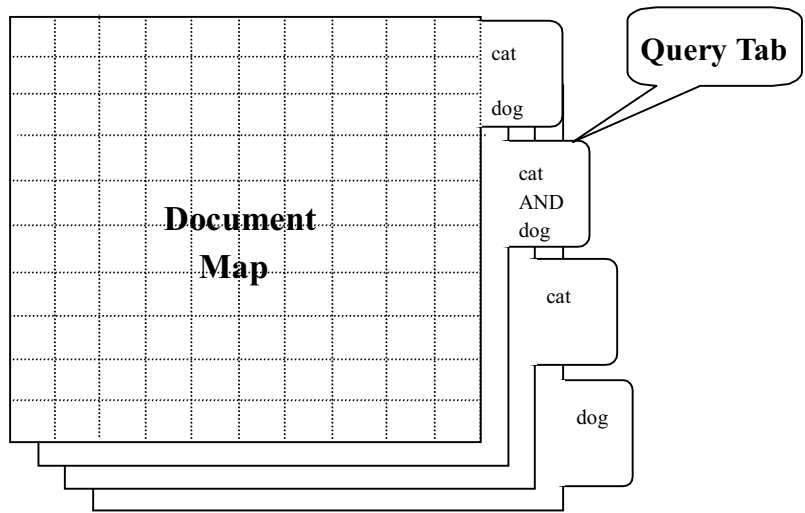

Fig. 2 Maps with query term tabs

Fig. 2 and Fig. 3 shows the basic concept of our visual display. The layers of document maps correspond to different rankings or clusters, respectively, in the figures, and each cell in a map represents a document. For example, the first map at the front in Fig. 2 represents the degree to which the 100 retrieved documents[1] are similar to the query vector (cat and doc). Those behind the first one indicate the degree to which the documents contain the Boolean combinations of 'cat AND dog', 'cat', and 'dog', respectively.

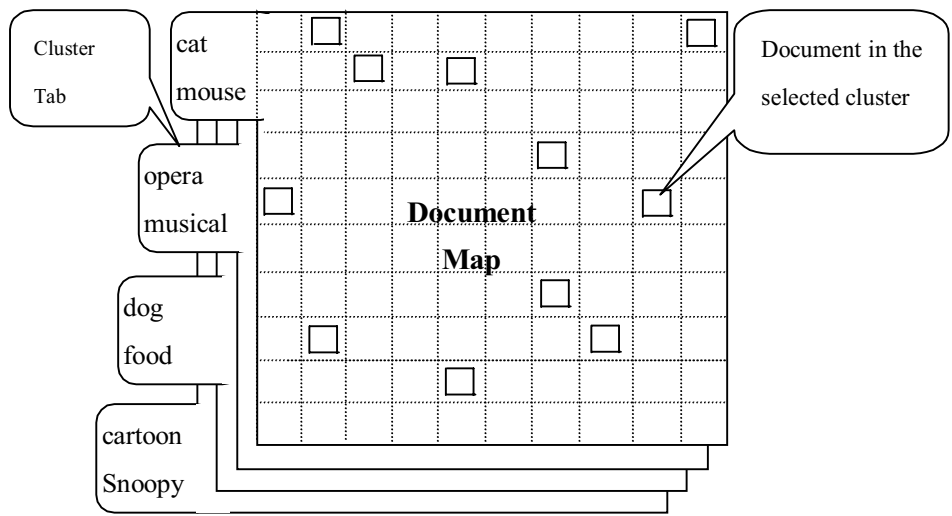

Fig. 3. Maps with cluster taps

The maps in Fig. 3 correspond to the clusters whose contents are represented by two words on the taps. On each map, those documents belonging to the cluster are marked appropriately. While the quality of the clusters is important for the efficacy of the visualization in our work, it is also important to ensure the clustering algorithm in use is efficient enough to handle the amount of the data contained in the documents at hand. For the current work, we simply adopted the algorithm devised to cluster web documents [16] primarily because of its efficiency. It should be noted that one query term tap and one cluster tap can remain selected to show two different views at the same time.

---

[1] It is possible to show more than 100 documents on a map as long as cells are big enough to be chosen easily by a mouse click.

Although not shown in the figures, the brightness of the color for each cell corresponds to the degree to which the document contains the word in the tab or to the cluster represented by the words in the tab. By selecting a tab, the user can switch to the particular view, i.e., ranking and/or cluster. While the user can only see one ranking with one specific cluster at a given point in time, he or she should be able to make a good guess about the content of a document by selecting different taps dynamically.

There are several features in this design that would help minimize possible confusion with many different views or maps. First, each map as well as the corresponding tab has its own unique color. Second, the cells in the same position across the layers of maps refer to the same document. In this way, the user does not have to remember where the document is on different maps; he or she can only see different brightness and color by switching among maps. Third, after a cell is clicked on, it appears to be pressed until it is clicked again so that the examined document remains to be shown as such even when different maps are chosen afterwards.

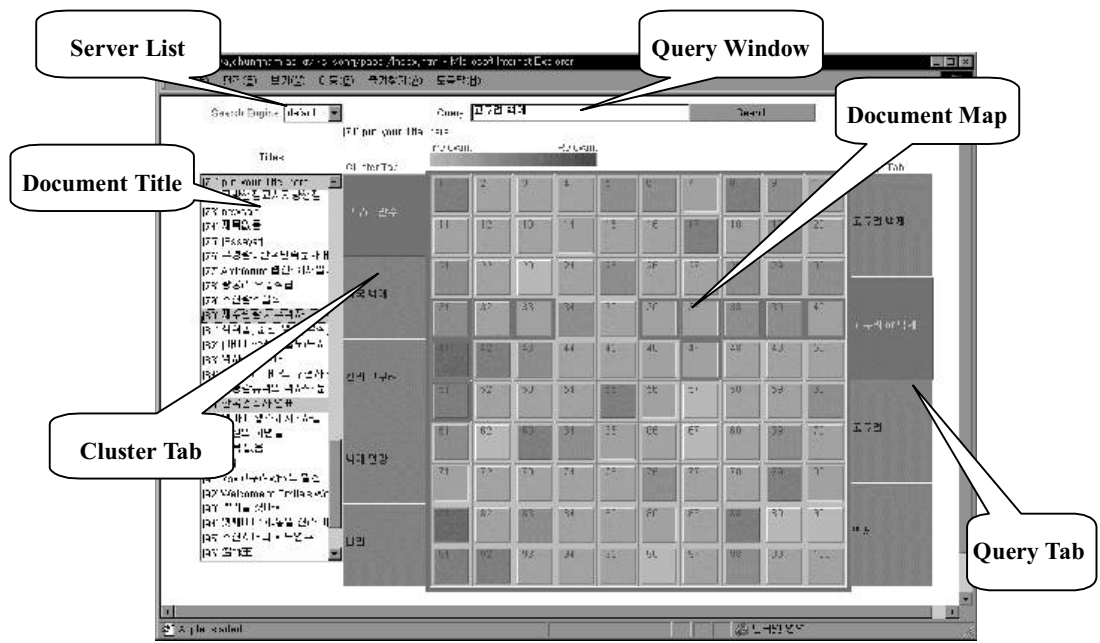

Fig. 4. An example of the document map interface

Fig. 4 shows an actual display of our document map visualization. At the upper left corner is the window for selecting a retrieval engine for which an adapter is available. Next to that is the query window in which users can type in a list of words. Below the engine selection window, a list of retrieved document titles corresponding to the document map is displayed when the retrieval engine returns the result. The current map displays which documents belong to the first cluster and to what extent they contain the two query terms together. The titles for the documents that have been examined by the user have been shaded in the list and the corresponding cells pressed. The color bar above the document map shows the brightness scale between the minimum and maximum similarity values. An important feature is that the pressed cells and the shaded titles remain the same even when different tabs are selected.

## 4. Experiments

While we felt our successful implementation of the presentation server and the client showed that the idea is viable for a user interface that requires heavy computation, we proceeded to test the efficacy of the particular visualization method so that it could be used for our MIRAGE DL system [9]. The goal of the experiments was to see how helpful the visualization method would be in users' task of identifying the relevant documents from the set of retrieved documents.

In order to see if the visual interface actually helps users select more relevant documents before irrelevant ones, we compared two retrieval situations: one only with the original list of documents and the other with the visual interface. Five human subjects, all graduate students, were asked to retrieve 100 documents from a given retrieval server and use the visual interface in any way they can to select the documents in the order they want based on the perceived relevance. As a result, they each created a new list of documents for each query, which was compared against the original list provided by the retrieval server.

We considered 30 pairs of lists as two sets of retrieval results and measured precision and recall. While indirect, Table 1 shows that the visual interface actually helps bringing up relevant documents to the front of the list when used appropriately. Fig. 5 shows the achieved precision after a certain number of documents have been reviewed. This indicates that users can get the desired number of relevant documents more quickly with the help of the visual interface. After reviewing 10 documents, for example, about 3.5 documents were relevant when the visual interface was used whereas only 1.5 relevant documents were included in the original list. In other words, users would be more satisfied with the particular order by which they examine the documents using the visual interface than with the original order.

Table 1. Comparison between the original list and the new list created with the visual interface

| Recall Level | Original List | With Document Maps | % Increase |
|---|---|---|---|
| 0.0 | 0.4502 | 0.6632 | **47** |
| 0.1 | 0.3405 | 0.5659 | **66** |
| 0.2 | 0.2633 | 0.4683 | **78** |
| 0.3 | 0.2390 | 0.4307 | **80** |
| 0.4 | 0.2223 | 0.3847 | **73** |
| 0.5 | 0.2058 | 0.3611 | **75** |
| 0.6 | 0.1948 | 0.2915 | **50** |
| 0.7 | 0.1884 | 0.2512 | **33** |
| 0.8 | 0.1840 | 0.2325 | **26** |
| 0.9 | 0.1776 | 0.2061 | **16** |
| 1.0 | 0.1750 | 0.2033 | **16** |
| **Average** | **0.2401** | **0.3690** | **51** |

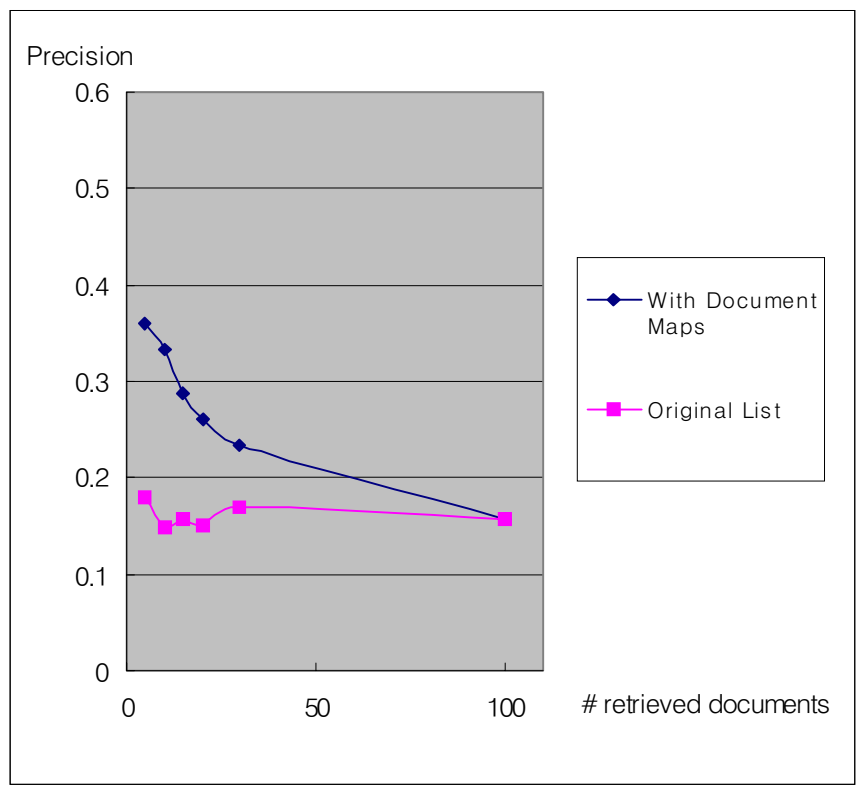

Fig. 5. Comparison in terms of precision values at various cutoff points

Normalized recall values for individual queries were calculated to show a somewhat different perspective from the experimental data. The measure computes the area difference between two curves representing progressive recall values obtained as new documents are considered one by one. On the average over the 30 queries, the normalized recall values for the original list and the new list of 100 documents are 0.5325 and 0.6624, respectively, yielding 24% improvement. This percent increase can be interpreted as the amount of reduction of users' effort in reviewing documents to get the same number of relevant documents. It should be noted that the percent increase gets larger and more significant as the number of documents considered gets smaller.

## 5. Conclusion

The idea of placing an intermediary between servers and clients has been found in other application areas. For example, an intermediary called Component Server plays an useful role of reducing burdens of the client in an object-oriented DBMS environment [10]. In a geographic information system a server called IOU takes the burden of processing image data for clients [6]. We believe, however, that our attempt to build a presentation server for search output in a DL environment is unique in that it facilitates visualization of search output, which has a direct bearing on users' ability to obtain relevant documents quickly.

With the presentation server, we can build a thin client by reducing the burden of processing a large amount of data for the purpose of displaying sophisticated visualization. Another advantage is that individual clients do not have to understand different types of search results coming from a variety of retrieval servers. While the heavy computation can be done in retrieval servers, it is also inefficient for them to deal with potentially many different types of data preparations for a variety of visualization requirements.

In this context, we also devised our own visualization interface in order to show that the presentation server indeed makes information visualization more practical. This interface requires computations for clustering of search output and for re-calculations of document similarities with respect to different sub-queries. By allowing users to view a set of retrieved documents from different perspectives through layers of document maps, the interface helps users in selecting desired documents more quickly. This functionality is particularly important when it takes time to bring in the full document descriptions from a remote server. Without the existence of a presentation server, this kind of visual interface may not be very practical.

Finally we ran experiments to demonstrate the usefulness of the visual interface. With the 10 by 10 document maps, which can be expanded if necessary, we obtained quite promising results. It was clear that users get more relevant documents early without having to review too many irrelevant documents when the interface is used.

The presentation server and the visual interface have been implemented as part of our ongoing MIRAGE project where new technologies are developed and integrated as a testbed for the ultimate purpose of transferring the technologies to the industry. We plan to refine the visual interface further and develop a variety of adapters so that the techniques can be actually used for real digital library environments. At the same time, we are working on an agent-based architecture where the presentation server behaves as an agent.

## Reference

[1] Benford, S. et al. (1995). "VR-VIBE: A Virtual Environment for Co-operative Information Retrieval", Proc. of Eurographics '95, Maastricht, The Netherlands.

[2] Cutting, D. R. et al. (1992). "Scatter/Gather: A Cluster-Based Approach to Browsing Large Document Collections," Proc. of the 15$^{th}$ ACM SIGIR Conference.

[3] Han, K. A. & Myaeng, S. H. (1996). "Image Organization and Retrieval with Automatically Constructed Feature Vectors." Proc. of the 19th International Conference on Research and Development in Information Retrieval, Zurich, Switzerland, August.

[4] Hearst, M. A. (1994). "Context and Structures in Automated Full-Text Information Access," Tech. Report UCB:CSD-94-836, Computer Science Department, Univ. of California, Berkeley.

[5] Heath, L. S. et al. (1995). "Envision: A User-Centered Database of Computer Science Literature," Communications of the ACM, 38 (4).

[6] ION (IDL On the Net): http://www.rsinc.com/ION/index.html

[7] Java Plug-in: Resource Page. http://java.sun.com/products/plugin/

[8] Lin, Xia (1996). Graphical Table of Contents, Proc. of the 1$^{st}$ International Conference on

Digital Libraries, Bethesda, MD, March.

[9] Myaeng, S. H. (1998). "R&D for a Nationwide General-Purpose System," Communications of the ACM, 41 (4), April.

[10] ObjectStore Component Server : http://www.datec.co.kr/tech/white/csf.html

[11] Olsen, K. A. et al. (1993). "Visualization of a Document Collection: The VIBE System," Information Processing and Management, 29 (1).

[12] RMI (Remote Method Invocation): Resource Page. http://java.sun.com/docs/books/tutorial/rmi/

[13] Robertson, G. G. et al. (1991). "Cone trees: Animated 3-D Visualizations of Hierarchical Information," Proc. of the ACM SIGCHI Conference on Human Factors in Computing Systems, April.

[14] Silverstein, C. & Pederson, J. (1997). "Almost Constant-Time Clustering of Arbitrary Corpus Subsets," Proc. of the 20[th] ACM SIGIR Conference, Philadelphia, PA.

[15] Young, P. (1996). "Three Dimensional Information Visualization," Tech. Report 12-96, Visualization Research Group Center for Software Maintenance, Dept. of Computer Science, University of Durham.

[16] Zamir, O. & Etzioni, O. (1998). "Web Document Clustering: A Feasibility Demonstration," Proc. of ACM SIGIR, Melbourne, Australia.